\documentclass[10pt]{article}
\usepackage[OE]{express}
\usepackage{subfig}
\usepackage{hyperref}
\usepackage[export]{adjustbox}
\usepackage{subfig}
\begin{document}

\title{All-Optical Nanopositioning of High-Q Silica Microspheres}
\author{Rafino M. J. Murphy,\authormark{1} Fuchuan Lei,\authormark{1*} Jonathan M. Ward,\authormark{1} Yong Yang,\authormark{1} and S\'ile Nic Chormaic\authormark{1}}
\address{\authormark{1}Light-Matter Interactions Unit, Okinawa Institute of Science and Technology Graduate University,
Onna, Okinawa 904-0495, Japan.\\
}
\email{\authormark{*}fuchuan.lei@oist.jp} %% email address is required

% // ----------------------------- abstract and OCIS codes ---------------------------- //
%
\begin{abstract}
A tunable, all-optical, coupling method has been realised for a high-\textit{Q} silica microsphere and an optical waveguide. 
By means of a novel optical nanopositioning method, induced thermal expansion of an asymmetric microsphere stem for laser powers up to 171~mW has been observed and used to fine tune the microsphere-waveguide coupling.  Microcavity displacements ranging from (0.612~$\pm$~0.13) -- (1.5~$\pm$~0.13) $\mu$m and nanometer scale sensitivities varying from (2.81~$\pm$~0.08) -- (7.39~$\pm$~0.17) nm/mW, with an apparent linear dependency of coupling distance on stem laser heating, were obtained. Using this method, the coupling was altered such that different coupling regimes could be explored for particular samples. This tunable coupling method, in principle, could be incorporated into lab-on-a-chip microresonator systems, photonic molecule systems, and other nanopositioning frameworks. 
\end{abstract}
\ocis{(070.5753) Resonators; (140.3948) Microcavity devices.}
%
% // ---------------------------------- References ----------------------------------- //

%\bibliographystyle{osajnl.bst}
%\bibliography{ref}

% // -----------------------------------  Intro  ------------------------------------- //
%
\section{Introduction}
\label{intro}
Whispering gallery mode (WGM) resonators have shown much promise in terms of their versatility and  scope in the last number of decades. These microcavities, with inherently small mode volumes $(V_m)$ and high-\textit{Q} factors, allow for strong light-matter interactions and have become widely used in bio-sensing and nanoparticle detection  \cite{Vollmer2008,Zhi2017,He2011a,Righini2016}, temperature/refractive index/pressure sensing  \cite{Ward2014,Yang2016a} quantum optics and electrodynamics  \cite{Matsko09,aoki2006}, microlaser development \cite{Yang2016b, Ward2016}, and have been used as a means of exploring optomechanical \cite{Lei15} and nonlinear effects \cite{Yang2016, xiao}. Despite  extensive research conducted on microcavities of various  geometries,  fabrication methods, and properties \cite{Sumetsky2010a,Tada2012,maayani2016water,liu2016fabrication}, the commonplace bulky experimental apparatus used in microresonator experiments has impeded their successful incorporation into many lab-on-a-chip or miniaturized systems \cite{Xu16}. 

When conducting experiments of this nature, it is imperative that the coupling between the microresonator and waveguide can be manipulated in such a way that a high coupling efficiency and low loss are maintained. It was first noted by Knight et al. \cite{K97} that a tapered silica fiber could be use to excite high-Q whispering gallery modes in silica microspheres. Later, Cai et al. \cite{Cai99} observed a critical coupling regime for the same experimental framework. The ease with which tapered fibers can be integrated into optical networks, as well as their inherent high coupling efficiency, are some of the many characteristics that make tapered fibers a favored option over other coupling methods (e.g. a prism). 

In order to further augment the sensitivity of these WGM systems post-fabrication, the coupling between a given resonator and optical source must be ameliorated. With regards to microsphere resonator systems, finely tuned coupling is usually achieved by means of a mechanical or piezoelectric nanometer scale positioner. Albeit useful, such  devices are difficult to incorporate into miniaturized microsphere or lab-on-a-chip systems that require a tunable coupling mechanism. Other non-mechanical means of realizing coupling regime control have been explored in the past \cite{Yang15}.  In this paper, we examine a means of achieving nanometer scale tunable coupling by taking advantage of thermo-mechanical effects arising from asymmetric microsphere stem fabrication, external laser heating, and thermal expansion in  single mode optical fiber. 

In principle, the scope of this work does not have to be restricted to microresonator coupled waveguide frameworks. With careful development and implementation this method could be used within photonic molecule systems as a means of varying the coupling between adjacent microcavities \cite{Boriskina2010,yang2017realization} and enhancing light-matter interactions, or perhaps as a micron scale nanostage, where the sphere acts as holder for objects which can then be positioned by simply varying the laser power.

% // --------------------------------- Fabrication  --------------------------------- //
%
\begin{figure}[b]
\centering
\includegraphics[width=10.5cm]{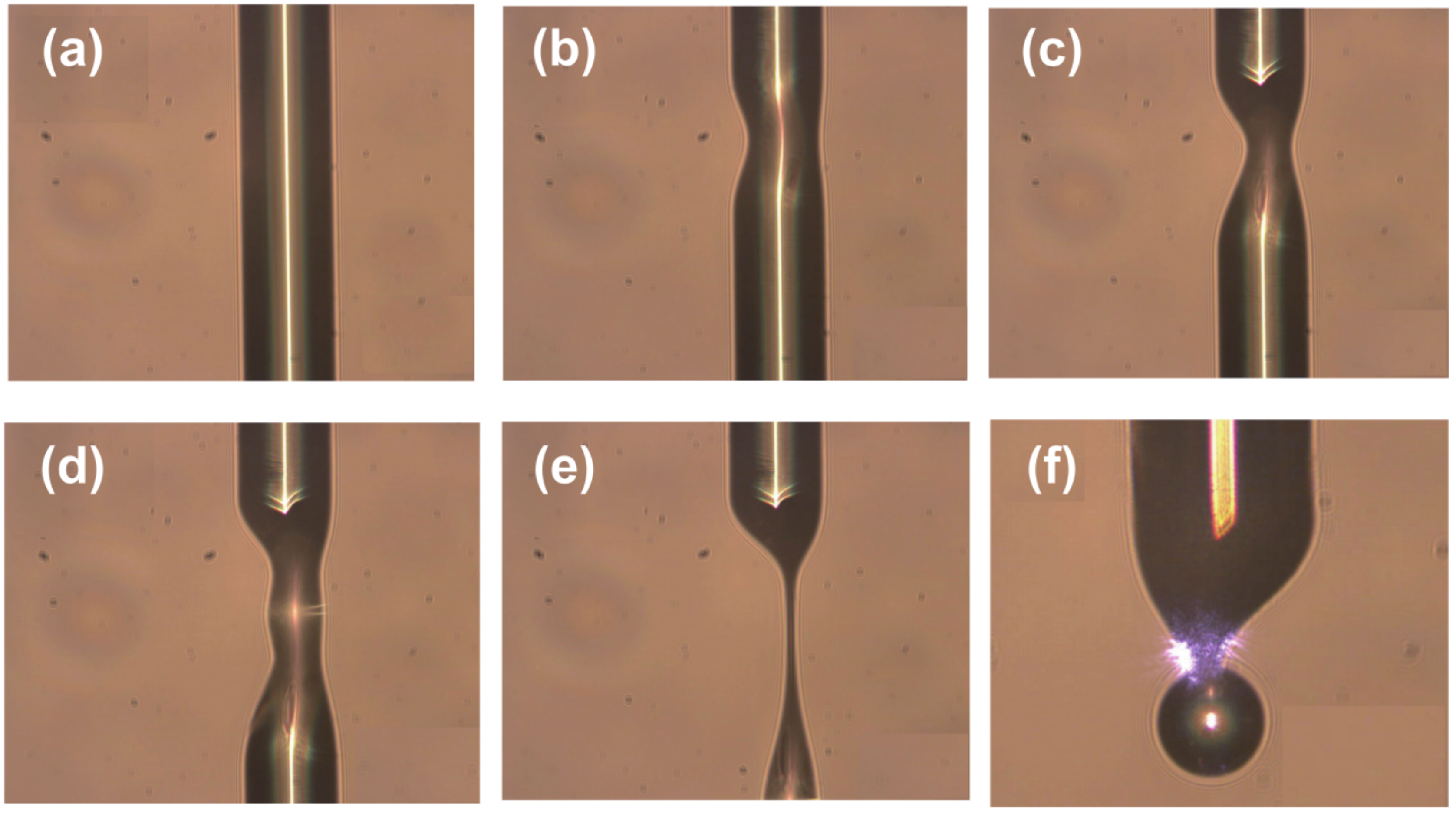}
\caption{\footnotesize{Asymmetric stem fabrication. \textbf{(a)} - \textbf{(c)}: The initial state of the optical fiber and the initial asymmetry resulting from side heating with a sub-fiber diameter CO$_2$ laser spot size and $\sim12$\% power. \textbf{(d)} - \textbf{(e)}: Resulting asymmetry after ablation with reduced laser spot size and decreased power (7-9\%). \textbf{(e)}: Final asymmetric-stem microsphere sample after careful ablation highlighting skewed core (center) and the external laser scattering region (illuminated by the 980 nm laser used in the experimental setup). A video of the fabrication process is also available (see supplementary information).}}
\label{fig:stem}
\end{figure}
\section{Fabrication Method}
\label{fab}
Initially, we made microspheres in the standard manner using a focused CO$_2$ beam (total power $\sim$25 W) directed onto a piece of silica single-mode optical fiber (Thorlabs SMF-28). A small weight attached to the bottom of the fiber upon heating ensured the formation of a tapered section which acts as the stem of the microsphere. By using a beam that was focused to a size spot smaller than the fiber diameter and adjusting the position of the beam such that the heating was predominantly on one side of the fiber, a slight asymmetry was formed during tapering. 
Generally, the initial asymmetry caused by side-heating of the optical fiber was counteracted by the downward pulling of the weight. In order to compensate for this, and to reinstate the stem asymmetry, the spot size was further reduced to the smallest spot size possible, and both the laser power (7-9\% power with the tight focus) and fiber position were manipulated to mimic an etching-like process via ablation -- essentially carving the asymmetry out of the fiber (see Fig.~\ref{fig:stem}). Conducting the ablation carefully causes the fiber core to taper towards one side (see Fig.1(b)), directing any incident light towards the asymmetric region. If the core is too straight, light will pass from the stem to the end of the microsphere, resulting in no scattering within the stem, no thermo-mechanical effects and no microsphere displacement. The next stage of the fabrication process involved defocussing the laser beam and heating the optical fiber away from the stem. The melted silica at the fiber tip assumed a spherical morphology, due to surface tension. Finally, we conducted further broad-focus laser heating of the microsphere to reduce surface irregularities. Post fabrication, the opposite end of the microsphere fiber was spliced to the output from a 980 nm laser diode source. The 980 nm laser was a standard EDFA pump type laser with a linewidth of a few nanometers and a maximum output power of ~211 mW. \vspace{-0.5ex}
% // --------------------------------- Asymmetry  --------------------------------- //
%
\begin{figure}[t!]
	\centering
	\subfloat[]{{\includegraphics[width=0.55\columnwidth]{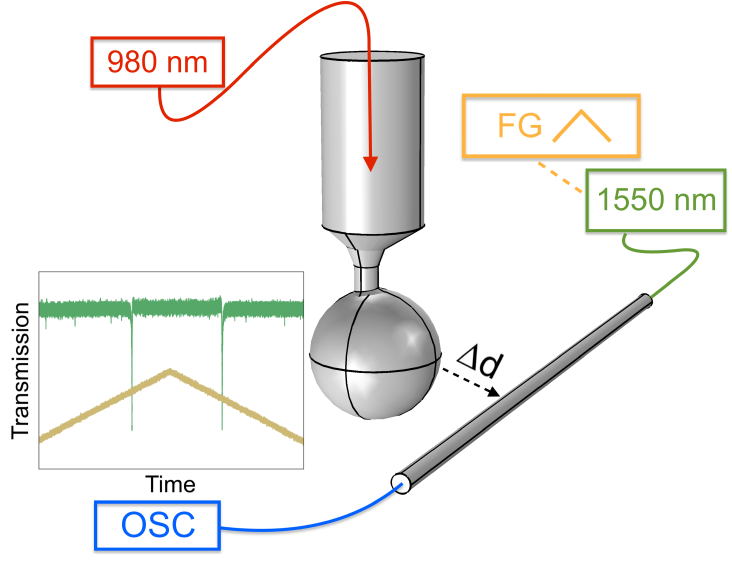}}}%
    \subfloat[]{{\includegraphics[width=0.325\columnwidth]{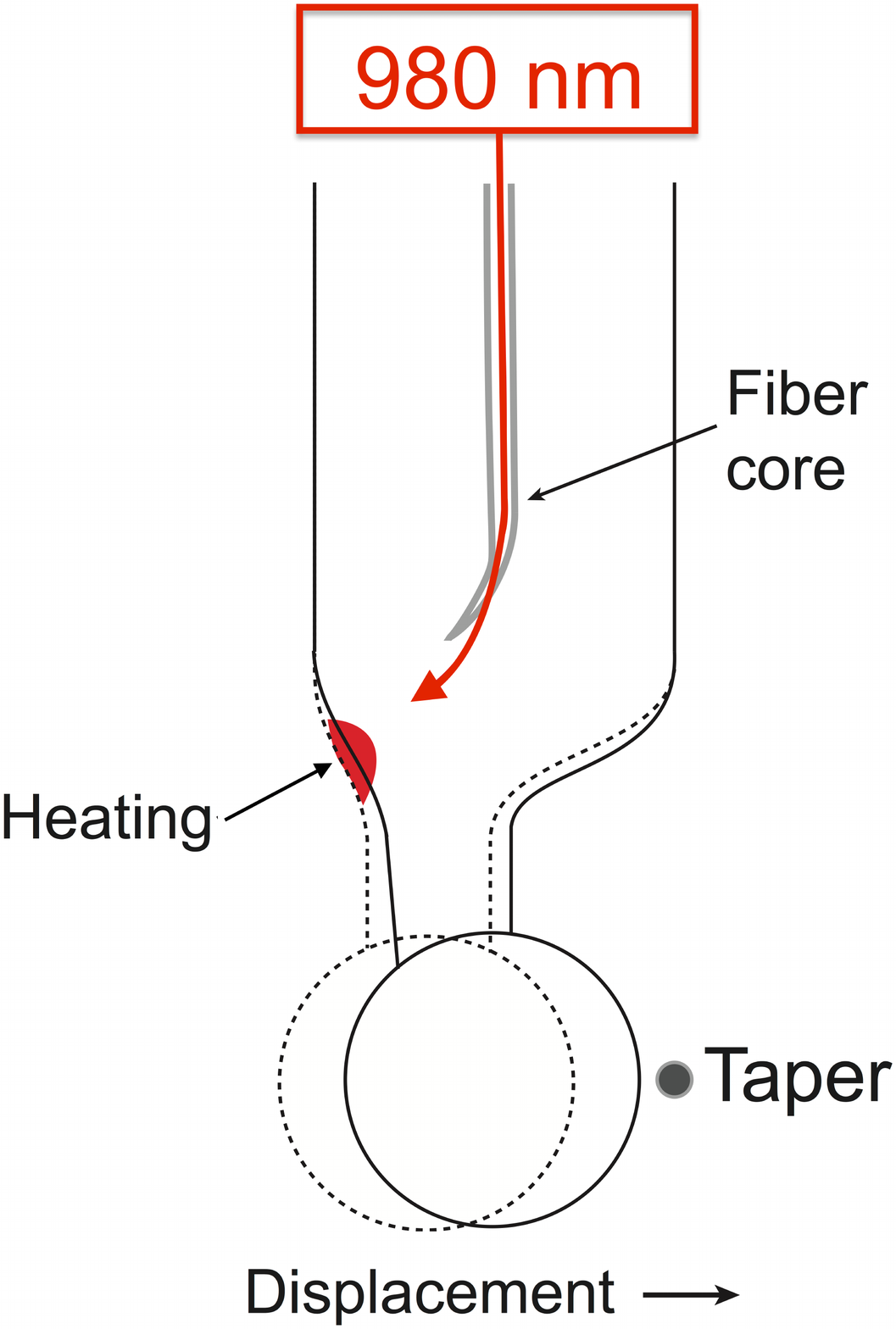}}}%
    \caption{\footnotesize{\textbf{(a)} Schematic diagram of the experimental set-up used to characterize the tunable thermo-mechanical coupling. OSC: Oscilloscope; FG: Function generator. \textbf{(b)} The fiber core during the fabrication process comes skewed towards the end, redirecting the incident laser light towards the asymmetric stem. The local surface conditions of the asymmetric stem region, once exposed to the external 980 nm laser source, exhibit localized heating and thermal expansion of the silica (red: an example of a scattering/heating region), leading to an increase (or decrease) in the coupling distance, $\Delta$$d$. For the sake of clarity, $\Delta$$d$ here is the distance between the microsphere and the tapered fiber.\vspace{-3ex}}}
    \label{fig:schema}	
\end{figure}
\section{Asymmetry and heating effects}
\label{asymm}
Normally, when light passes from the end of an optical fiber to the surrounding medium there are minimal thermal effects. The light traversing the dielectric material passes through a predominantly uniform cross-sectional area, $a$, into the surrounding medium with no resultant heating of the silica. In our samples, the fiber cores have been skewed to one side during the fabrication process (see Section \ref{fab}), altering the direction of propagation of light within the fiber and directing it towards their respective asymmetric stem regions (see Fig.~\ref{fig:stem}). The geometry and non-uniformity of the asymmetric stem regions appear to prevent the laser light from passing unimpeded from the symmetric cylindrical region of the optical fiber to the end of the microsphere. Instead, the light appears to undergo scattering at different regions of the asymmetric domain depending on the fiber core orientation.

The effects of laser energy deposition and laser damage on imperfect/deformed dielectric surfaces has been thoroughly examined over the last few decades \cite{Demos2013,Bloembergen1973}. The aforementioned research suggests that these imperfections -- such as nanometer to micrometer-scale cracks or spherical pores -- can alter the amount of energy deposited at the laser-dielectric interaction boundary. The preceeding experimental results have led us to conclude that the orientation of the fiber core and geometry of the asymmetric stem invoke local surface conditions that induce scattering of the light within the stem, resulting in uneven energy deposition throughout the silica, and localized heating and thermal expansion. The net thermal expansion manifests as a linear displacement of the microsphere, such that, when placed in a WGM coupling  set-up, an increase or decrease in tapered fiber transmission is observed, depending on the orientation of the microsphere.
%
%
%
% // --------------------------------- Procedure & results  --------------------------------- //
%
\section{Tunable Thermo-Mechanical Coupling Characterization}
\begin{figure}[b!]
	\centering
 	{\includegraphics[width=0.9\columnwidth]{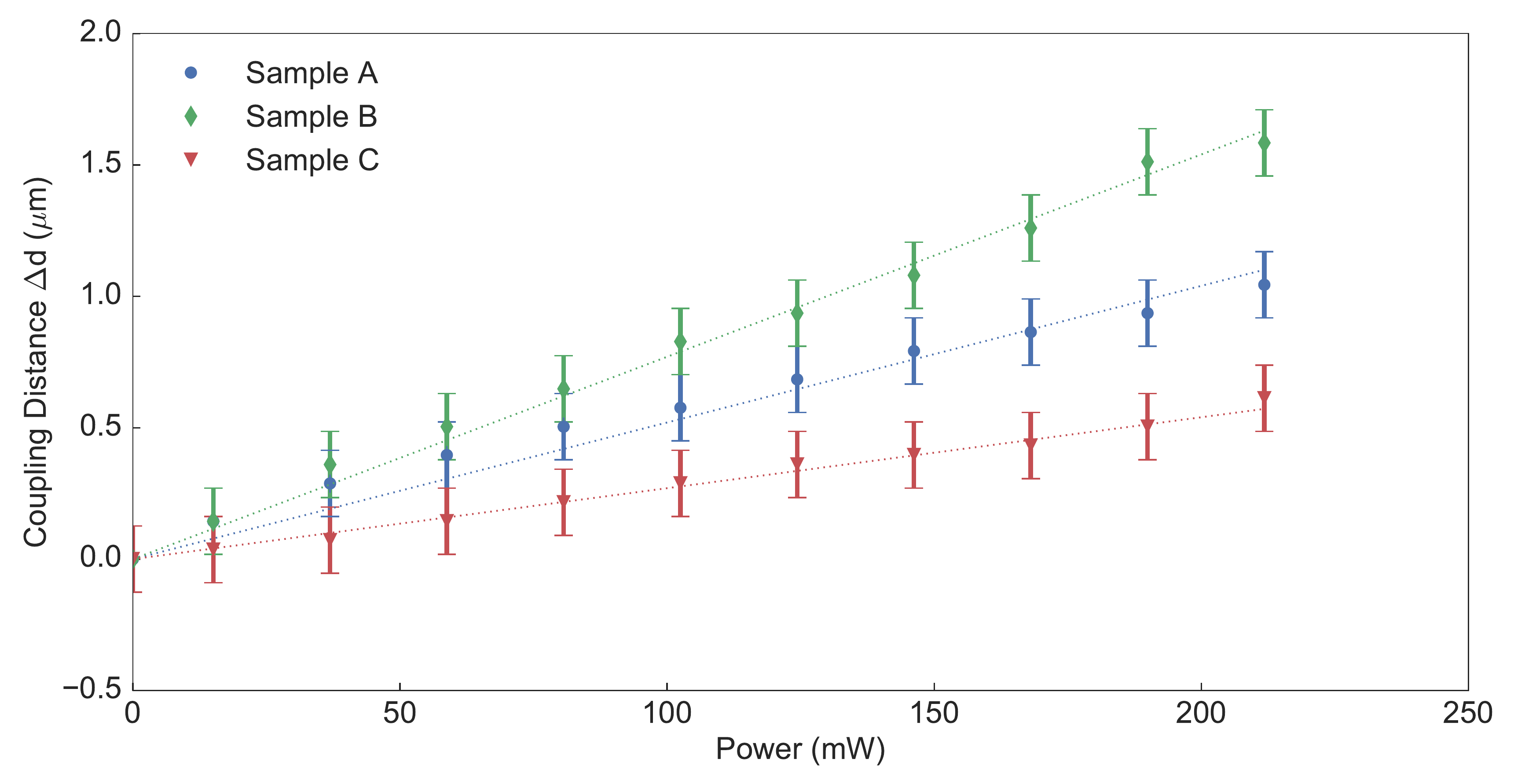}}%
	\caption{\footnotesize{Graph of coupling distance, $\Delta d$, against laser power for three samples A, B and C. The results indicate a linear increase in coupling distance with increasing laser power for these microsphere orientations, consistent with the thermal expansion hypothesis when the orientation and scattering regions are taken in to consideration. Displacement sensitivities of (7.39~$\pm$~0.17), (4.42~$\pm$~0.12) and (2.81~$\pm$~0.08) nm/mW and a total $\Delta$$d$ of 1.58~$\mu$m, 1.04~$\mu$m and 0.61~$\mu$m are observed.}}
	\label{fig:dist}
\end{figure}
%\subsection{Coupling Distance} 
%\textbf{(i) Coupling Distance:} 
In order to isolate the aforementioned microsphere displacement and ascertain how it affects coupling efficiency, we used a standard WGM resonator-tapered fiber set-up  (see Fig. ~\ref{fig:schema}(a)). As a means of checking the validity of our hypothesis, the location of the scattering region was examined first, after which the microsphere was positioned such that the asymmetry was present in the plane perpendicular to the tapered fiber. Initially, static forces and other factors which may directly affect coupling over time were mitigated by isolating the system for at least four or five hours, and ensuring that, prior to the experiment, sufficient tension was applied to the tapered fiber post-fabrication. Using a piezoelectric nano-positioning stage (Thorlabs 3-Axis Piezo Controller MDT693A), the microsphere was brought close to the tapered fiber without initiating contact. 

The coupling depth of the selected mode in the observed transmission spectrum and the piezo voltage ($V_c$) were used as initial position references. The laser power was then incremented from 0 to 211 mW, causing the coupling depth of the chosen mode in the transmission spectrum to change. Then, by means of the piezoelectric nano-positioning stage, the microsphere was moved until the initial coupling depth was reinstated. The voltage difference applied to the piezo stage between this new position ($V_o$) and the initial reference position ($V_c$) was used to determine the change in coupling distance ($\Delta d = |V_c - V_o| \times k_p$), where the measured sensitivity ($k_p$) of the piezo stage was 0.36 $\mu$m/V and the rated resolution of the piezo stage is 27 nm. In order to exclude the possibility of cavity heating causing the observed transmission spectrum change, the relative change in transmission over the whole laser power range was determined for each sample. Also, as an extra means of precaution, each sample's non-contact transmission spectrum was obtained then isolated for a period of time. If the coupling did not vary over time, the subsequent results that were taken were deemed valid.

\begin{figure}[h!]
	\centering
 	{\vspace{-0.2ex}\includegraphics[width=0.9\columnwidth]{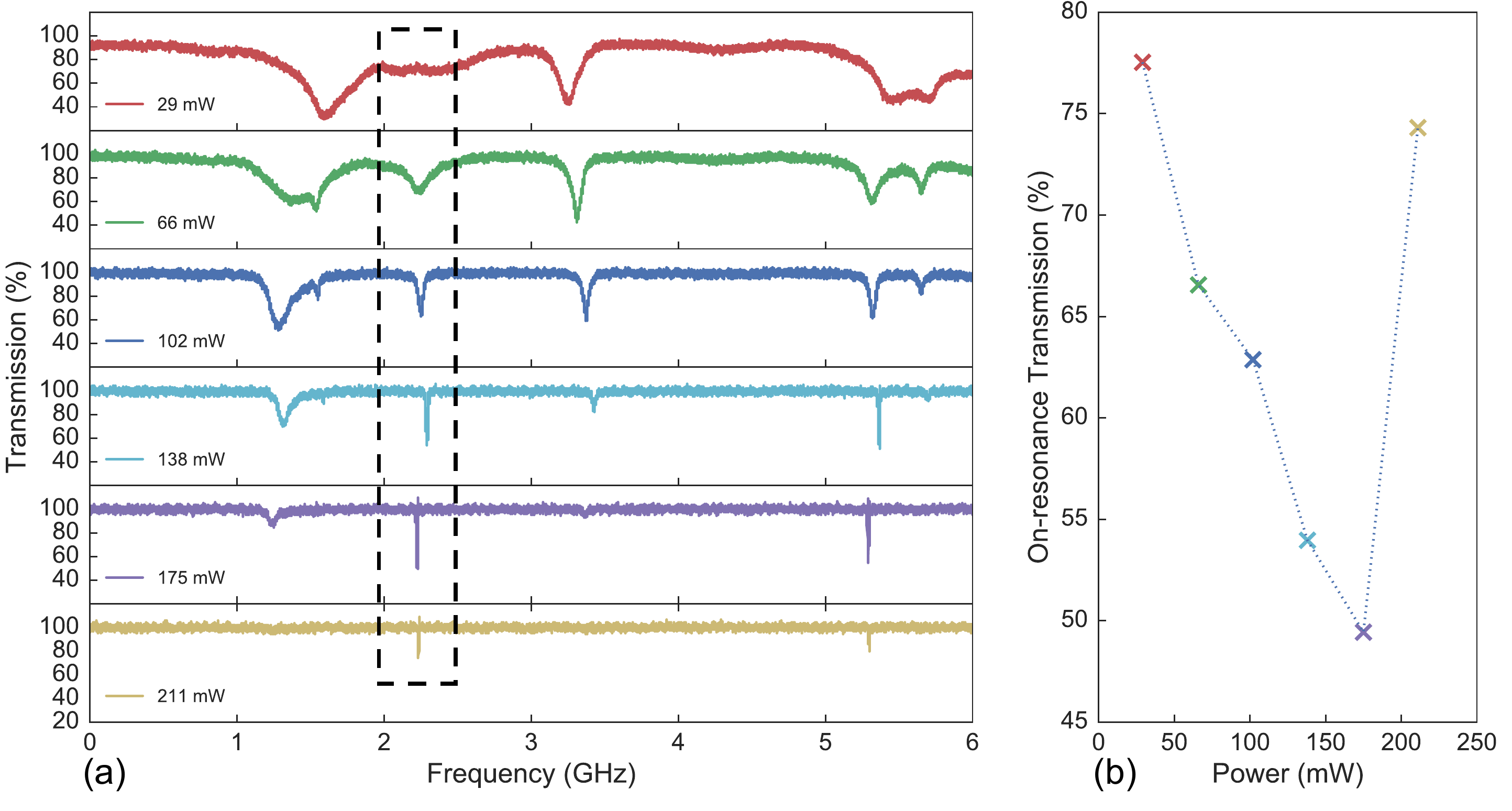}}%
	\caption{\footnotesize{\textbf{(a)} Sample D. Series of normalized transmission spectra taken at a laser scanning wavelength of 1540 nm for increasing laser power from 29 -- 211 mW. Over this range, a total mean transmission change of 23\% is observed. The spectra have been normalized to the 211 mW transmission spectrum. \textbf{(b)} Percentage transmission of a whispering gallery mode (dashed enclosure, see Fig. \ref{fig:qval}(a)) over the aforementioned laser power range. This particular whispering gallery mode initially exhibits over-coupled behavior at a laser power of 29 mW. Increasing laser power results in an increase in coupling distance and a decrease in on-resonance transmission until the mode exhibits (low-\textit{Q} equivalent) ``critical'' coupling at a laser power of 175 mW and a total transmission of 49\%. At a maximum laser power of 211 mW, the mode enters the under-coupled regime and on-resonance transmission increases to 74\%.\vspace{-4ex}}}
	\label{fig:qval}
\end{figure}

%\results{Coupling Distance} 
The change in coupling distance, $\Delta d$, for three independent samples, A, B and C, is shown in Fig. ~\ref{fig:dist}. The results suggest an increase in coupling distance with increasing laser power. By checking the orientation of the microsphere and examining the scattering region of the light in each case (see Fig. \ref{fab}(f)), the results are consistent with the thermal deformation/stem thermal expansion hypothesis. Sample A, B and C exhibited displacement sensitivities of (7.39~$\pm$~0.17), (4.42~$\pm$~0.12) and (2.81~$\pm$~0.08) nm/mW and total displacements $\Delta$$d$ of 1.58~$\mu$m, 1.04~$\mu$m and 0.61~$\mu$m, respectively. The  transmission spectra at different input powers of the external laser, for a particular sample, can be seen in Fig.~\ref{fig:qval}(a).

%\noindent{\textbf{(ii) Coupling regimes:}} 
Whether or not the thermal deformation of the stem could be used to traverse the different coupling regimes (under, critical and over-coupling) was also examined. For a particular mode, in  sample D, the laser current was increased from 29 -- 211 mW (for a total power range of $\sim$182 ~mW) while the transmission spectrum for each power increment was recorded.
%\Results{Coupling regimes}
Examination of a particular mode with laser power variation over $\sim$182 mW shows evidence of traversing the different coupling regimes (see Fig. ~\ref{fig:qval}(b)). Initially, the microsphere is at a distance, $\Delta$$d$, such that the mode exhibits over-coupled behavior with the on-resonance transmission at 78\% (where the off-resonance transmission is defined as 100\%). As the laser power is increased and $\Delta$$d$ increases, the selected mode shows low-\textit{Q} and over-coupled behavior. As the pump power is increased the coupling of the mode increases along with the \textit{Q}  until at a power of 175 mW, the mode reaches the ``critical'' coupling regime with on-resonance transmission of 49\%. Strictly speaking, this behavior cannot be truly classified as critical coupling as the transmission does not approach 0\% resulting from a low \textit{Q}-factor ascribable to microsphere surface irregularities such as surface aberrations, dirt, input light polarization, etc. However, the evident behavior of this particular mode is equivalent to low-\textit{Q} critical coupling. As the laser power reaches a maximum and the coupling distance between the microsphere and tapered fiber waveguide increases, the mode enters an under-coupled regime with the on-resonance transmission increasing to 74\%.

% // --------------------------------- Conclusion  --------------------------------- //
\section{Conclusion}
\vspace{-0.75ex}
In conclusion, we have used an external laser and asymmetric-stem microsphere system to develop a tunable, thermo-mechanical coupling method within a WGM resonator-tapered fiber waveguide coupling system. Examination of this optical nanopositioning device with three independent samples shows a linear dependency of coupling distance with laser power, with sensitivities ranging from (2.81~$\pm$~0.08) -- (7.39~$\pm$~0.17) nm/mW. Under the right local conditions within the microsphere stem, traversing the different coupling regimes using this method was possible. In the future, this method could be used as a means of enabling tunable microsphere coupling in miniaturized or lab-on-a-chip resonator systems, photonic molecule systems and other nanopositioning systems.\vspace{-0.75ex}
%
% // --------------------------------- End  --------------------------------- //
%
\section*{Acknowledgments}
This work was supported by the Okinawa Institute of Science and Technology Graduate University.
\end{document}